\documentclass[prl,tightlines,twocolumn,seceqnum,showpacs,superscriptaddress,floatfix]{revtex4-1}
\usepackage[dvips]{graphicx}
\usepackage[english]{babel}
\usepackage{color,enumerate,amsmath,amssymb,amsbsy,amscd,bm}
\usepackage{epsf,epsfig}
\usepackage{tikz}
\usetikzlibrary{shapes.arrows}
\usetikzlibrary{arrows,decorations.pathmorphing}

\newcommand{\ket}[1]{|{#1} \rangle}  

\begin{document}
\preprint{}
\title[]{How Much is the Efficiency of Solar Cells Enhanced by Quantum Coherence?}
\author{Sangchul Oh}
\affiliation{Qatar Environment and Energy Research Institute,
Hamad Bin Khalifa University, Qatar Foundation, P.O. Box 5825, Doha, Qatar}
\date{\today}
\begin{abstract}
We study how much the efficiency of a solar cell as a quantum heat engine could be enhanced by quantum coherence.
In contrast to the conventional approach that a quantum heat engine is in thermal equilibrium with both 
hot and cold reservoirs, we propose a new description that the quantum heat engine is in the cold reservoir 
and the thermal radiation from the hot reservoir is described by the pumping term in the master equation.
This pumping term solves the problem of the incorrect mean photon number of the hot reservoir assumed 
by the previous studies. By solving the master equation, we obtain the current-voltage and the power-voltage curves 
of the photocell for different pumping rates. We find that, as the photon flux increases, the power output 
of the photocell increases linearly at first and then becomes saturated, but the efficiency decreases rapidly. 
It is demonstrated that while the power output is enhanced significantly by the quantum coherence via the dark 
state of the coupled donors, the improvement of the efficiency is not significant.
\end{abstract}
\maketitle

%

Solar cells and photosynthesis, which convert sunlight into electrical and chemical energies, 
respectively, may be regarded as heat engines. 
The maximum efficiency of a heat engine
operating between hot and cold reservoirs is known as the Carnot efficiency, derived from 
the second law of thermodynamics. For a quantum heat engine, Scovil and Schulz-DuBois considered 
a three-level maser in thermal contact with two heat reservoirs, and  showed its ultimate efficiency 
is that of a Carnot engine~\cite{Scovil59}. Shockley and Queisser obtained the upper limit of efficiency of a single 
{\it p-n} junction solar cell, based on the assumption that electron-hole pairs recombine only through 
the radiative process, i.e., the principle of the detailed balance~\cite{Shockley1961}. 
The Shockley-Queisser limit, however, is far below the Carnot efficiency because of 
only one electron-hole pair generation per photon with energy larger than 
the band gap of the semiconductor generates. 


Recent studies have shown that quantum effects could play a key role in photosynthesis and solar cells.
Engel and his co-workers observed the long-lived quantum coherence in exciton dynamics in the 
Fenna-Matthews-Olsen complex, using 2-dimensional electronic spectroscopy~\cite{Engel2007}. Following 
experimental and theoretical studies suggest that this quantum beat may be due to the interplay of 
electronic and vibronic quantum dynamics. Scully and his colleagues showed theoretically that quantum 
coherence could enhance the efficiency of a solar cell and a photosynthetic reaction
center~\cite{Scully2010,Scully2011,Svidzinsky2011,Dorfman2013}. 
It has been argued that the quantum coherence could break the detailed balance, and thus 
the Shockley-Queisser limit of the efficiency of solar cells. 

Inspired by Scully {\it et al.}'s work, Creatore {\it et al.}~\cite{Creatore2013} proposed a biologically 
inspired photocell model enhanced by a delocalized dark quantum state of two dipole-dipole coupled donors. 
Zhang {\it et al.}~\cite{Zhang2015} showed that the delocalized dark state of three coupled donors could 
enhance more the efficiency of a photocell.
Recently, Fruchtman {\it et al.}~\cite{Fruchtman2016} showed that a photocell with asymmetric pair of coupled chromophores 
could outperform those with the symmetric dimer or with a pair of independent molecules.

While theoretical studies on photocells mentioned above predict promising enhancement of the efficiency of a quantum
heat engine, there is controversy, especially, raised by Kirk~\cite{Kirk2010,Kirk2012,Kirk2015}.  
The claim of the role of quantum coherence in enhancing the efficiency needs to be more complete in the following sense.
First, photocells as a quantum heat engine are assumed to 
be in thermal equilibrium with hot and cold reservoirs simultaneously. This assumption may give rise to a question on 
the temperature of a photocell. Second, the average photon number of the Sun with a temperature of $6000\,{\rm K}$ 
at the energy gap of donors was incorrectly used in the master equations in previous studies. 
Finally, while the previous studies have shown the power enhancement by quantum effects, they tells neither 
how much efficiency is enhanced nor whether the Shockley-Queisser limit is surpassed.  

In the paper, we present a realistic model of a photocell which is in thermal contact only with the cold 
reservoir. The pumping term in a master equation is introduced in order to take into account the photon 
flux from the hot reservoir. This resolves the issue of the incorrect mean photon number of the hot reservoir 
assumed by the previous studies, and makes it possible to calculate the efficiency. The power output of 
the photocell is obtained as a function of the strength of the pumping term, i.e., the photon flux. We show 
that the power increases linearly at first but becomes saturated as the pumping strength increases. We obtain 
the efficiency as a function of pumping strength and demonstrate that quantum coherence could enhance the 
efficiency, but not much.

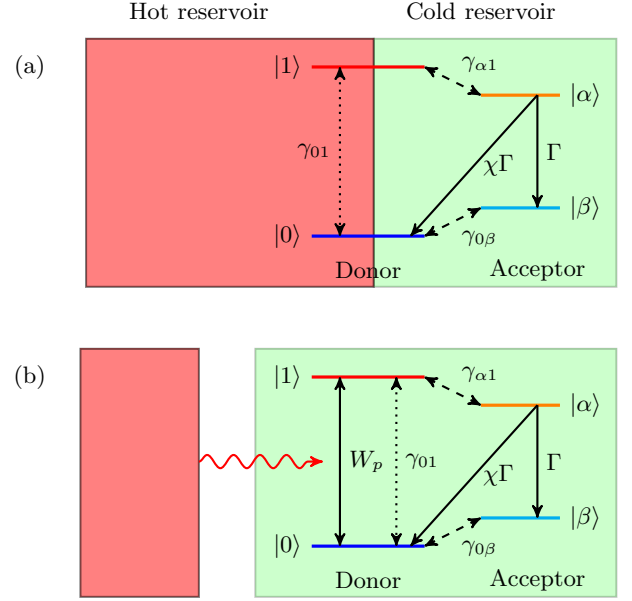
\begin{figure}[htbp]
\begin{tikzpicture}[thick, scale=0.75, 
>=stealth', pos=.8, photon/.style={decorate,decoration={snake,post length=1mm}} ]
\begin{scope}[shift={(0,5.5)}]
\draw[] (-5,3) node {(a)};
\draw[] (-2,4) node {Hot reservoir};
\draw[] ( 3,4) node {Cold reservoir};
\draw [fill=red, opacity=0.50] (-4.0,-0.9) rectangle  (1.1,3.5);
\draw [fill=green, opacity=0.20] (1.1,-0.9) rectangle  (5.4,3.5);
\draw[red, very thick] (0,3) node[left] {\color{black} $\ket{1}$} -- (2,3);
\draw[blue,very thick] (0,0) node[left]{\color{black} $\ket{0}$}  -- (2,0);
\draw[black,dotted,thick, <->]  (0.5,0) -- (0.5,3) node[midway, left] {$\gamma_{01}$};
\draw[black,dashed,thick, <->]  (2,3) -- (3,2.5) node[midway,above right] {$\gamma_{\alpha 1}$};
\draw[orange,very thick] (3,2.5) -- (4.4,2.5) node[right] {\color{black}$\ket{\alpha}$};
\draw[cyan,very thick]   (3,0.5) -- (4.4,0.5) node[right] {\color{black}$\ket{\beta}$};
\draw[black,thick, ->]  (4,2.5) -- (4,0.5) node[midway,right] {$\Gamma$};
\draw[black,thick, ->]  (4,2.5) -- (1.75,0.0) node[midway,right] {$\chi\Gamma$};
\draw[black,dashed,thick, <->]  (3,0.5) -- (2,0) node[midway,below right] {$\gamma_{0\beta}$};
\draw[] (1,-0.6) node {Donor};
\draw[] (4,-0.6) node {Acceptor};
\end{scope}
\begin{scope}[shift={(0,0)}]
\draw[] (-5,3) node {(b)};
\draw [fill=red, opacity=0.50] (-4.1,-0.9) rectangle  (-2,3.5);
\draw [fill=green, opacity=0.20] (-1,-0.9) rectangle  (5.4,3.5);
\draw[red, very thick] (0,3) node[left] {\color{black} $\ket{1}$} -- (2,3);
\draw[blue,very thick] (0,0) node[left]{\color{black} $\ket{0}$}  -- (2,0);
\draw[black,dotted,thick, <->]  (1.5,0) -- (1.5,3) node[midway, right] {$\gamma_{01}$};
\draw[black,thick, <->]  (0.5,0) -- (0.5,3) node[midway, right] {$W_p$};
\draw[color=red,->, photon]  (-2,1.5) -- (0.2,1.5);
\draw[black,dashed,thick, <->]  (2,3) -- (3,2.5) node[midway,above right] {$\gamma_{\alpha 1}$};
\draw[orange,very thick] (3,2.5) -- (4.4,2.5) node[right] {\color{black}$\ket{\alpha}$};
\draw[cyan,very thick]   (3,0.5) -- (4.4,0.5) node[right] {\color{black}$\ket{\beta}$};
\draw[black,thick, ->]  (4,2.5) -- (4,0.5) node[midway,right] {$\Gamma$};
\draw[black,thick, ->]  (4,2.5) -- (1.75,0.0) node[midway,right] {$\chi\Gamma$};
\draw[black,dashed,thick, <->]  (3,0.5) -- (2,0) node[midway,below right] {$\gamma_{0\beta}$};
\draw[] (1,-0.6) node {Donor};
\draw[] (4,-0.6) node {Acceptor};
\end{scope}
\end{tikzpicture}
\caption{(a) A photocell is in thermal equilibrium with both hot and cold reservoirs. (b) A photocell is in thermal
equilibrium only with the cold reservoir. The hot reservoir excites the donor with pumping rate $W_p$.}
\label{Fig1}
\end{figure}

\paragraph*{Solar Cell with Donor-Acceptor.---}
Let us start with a simple photovoltaic model, a four-level quantum system composed of a donor and a acceptor,
as shown in Figs.~\ref{Fig1} (a) and (b). We present the issue of the previous photocell models and solve it by
introducing the pumping term in our model. Fig.~\ref{Fig1} (a) depicts a photocell model of previous studies 
that is in the {\it thermal equilibrium with both hot and cold reservoirs at the same time}. The total Hamiltonian 
is written formally as 
\begin{align}
H = H_S + H_H + H_C + H_{SH} + H_{SC} \,,
\end{align}
where $H_S$ is the Hamiltonian of the photocell with donor and acceptor, and $H_H$ $(H_C)$ is the Hamiltonian of 
the hot (cold) reservoir represented by the collection of the harmonic oscillators. Typically, it is assumed that
the interactions, $H_{SH}$ and $H_{SC}$, between the system and the reservoirs are assumed to be time-independent. 
Using the Born and the Markov approximations, one can obtain the master equation for the system dynamics.

As shown in the previous
studies~\cite{Scully2011,Svidzinsky2011,Dorfman2013,Creatore2013,Zhang2015,Killoran2015,Fruchtman2016}, 
the probabilities $P_i$ of occupation of energy levels $E_i$ obey the Pauli master equations
\begin{subequations}
\label{Eq:4level_pv_old}
\begin{align}
&\begin{aligned}
\dot{P}_{0} &= \gamma_{01} \bigl[ (n^h_{01} +1)\,P_1  - n^h_{01}\,P_0 \bigr]+ \chi\Gamma\,P_\alpha\\[2pt]
            &+ \gamma_{0\beta}\bigl[ (n^c_{0\beta}+1)\,P_\beta -\,n^c_{0\beta}\,P_0 \bigr]\,,
\end{aligned}\\[5pt]
&\begin{aligned}
\dot{P}_{1} &= \gamma_{01} \bigl[ n^h_{01}\,P_0 - (n^h_{01}+1)\,P_1\bigr] \\[2pt]
            &+ \gamma_{\alpha 1}\bigl[ n^c_{\alpha 1}\,P_\alpha -(n^c_{\alpha 1}+1)\,P_1 \bigr] \,,
\end{aligned}\\[5pt]
&\dot{P}_{\alpha} =\gamma_{\alpha 1}\bigl[(n^c_{\alpha 1}+1)\,P_1 -n^c_{\alpha 1}\,P_\alpha\bigr] 
                  -(1+\chi)\Gamma\,P_\alpha\,,\\[5pt]
&\dot{P}_{\beta} = \gamma_{0\beta}\bigl[n^c_{0\beta}\,P_0 -(n^c_{0\beta}+1)\,P_\beta\bigr] + \Gamma\,P_\beta\,.
\end{align}
\label{PE1_old}
\end{subequations} 
Here $\gamma_{ij}$ are the transition rates between level $E_i$ to level $E_j$.
The mean photon number $n^{h}_{ij}$ ($n^{c}_{ij}$) of the hot (cold) reservoir at temperature $T_h$ $(T_c)$ for a given 
frequency $\Delta E_{ij} = E_j -E_i$ is written as
\begin{align}
n_{ij}^h = \frac{1}{e^{\Delta E_{ij}/k_BT_h} -1}\,.
\label{mean_photon}
\end{align}
The parameters are taken as follows: $E_1-E_0= 1.8\,{\rm eV}$, $E_1-E_\alpha = E_\beta -E_0 = 0.2 {\;\rm eV}$,
$\hbar\gamma_{01} = 1.24{\;\mu}{\rm eV}$, $\hbar\gamma_{\alpha 1} = 12 {\;\rm meV}$,
and $\hbar\gamma_{0\beta} = 24 {\;\rm meV}$
~\cite{Scully2011,Svidzinsky2011,Dorfman2013,Creatore2013,Zhang2015,Killoran2015,Fruchtman2016}.  
These imply $1/\gamma_{01}\simeq 0.5{\;\rm ns}$, $1/\gamma_{\alpha 1} \simeq 0.55{\;\rm fs}$, 
$1/\gamma_{0\alpha} \simeq 0.26\,{\rm fs}$, and $\chi = 0$.
So, the typical time to reach the steady state is the order of femotosecond.
The temperates of the hot and cold reservoirs are $T_h = 6000{\;\rm K}$ and $T_c = 300{\;\rm K}$, respectively.
If the parameters are plugged into Eq.~(\ref{mean_photon}), the mean photon number of the hot reservoir at energy
$\Delta E_{01}=1.8{\;\rm eV}$ is given by $n_{01}^h \simeq 0.0317$, and the mean photon number of the cold reservoir
at energy $\Delta E_{1\alpha} =\Delta E_{0\beta}= 0.2 {\;\rm eV}$ by $n^c_{1\alpha}=n^c_{0\beta}\simeq 4.368\times
10^{-4}$~\cite{Mandel}. However, the previous
papers~\cite{Scully2011,Svidzinsky2011,Dorfman2013,Creatore2013,Zhang2015,Killoran2015,Fruchtman2016} assumed $n^h_{01}
= 60000$ that does not coincide with the value given by Eq.~(\ref{mean_photon}).

In order to solve the pitfall of the previous studies depicted in Fig.~\ref{Fig1}~{(a)}, we propose a new 
photocell model as shown in Fig.~\ref{Fig1}~(b). The donor of the new photocell is assumed to be in 
thermal equilibrium only with the cold reservoir, but not with the hot reservoir. The photon flux from 
the hot reservoir is described by the pumping term~\cite{Slichter}. So the strength of the pumping term may 
correspond to the solar irradiance incident on the photocell. It is straightforward to obtain the Pauli master 
equations with the pumping term for the population dynamics of the new photocell model
\begin{subequations}
\label{PE1_new}
\begin{align}
&\begin{aligned}
\dot{P}_{0} 
&= \gamma_{01} \bigl[ ({\color{blue} n^c_{01}} +1)\,P_1  - {\color{blue} n^c_{01}}\,P_0 \bigr]+ \chi\Gamma\,P_\alpha\\[2pt]
&+ \gamma_{0\beta}\bigl[ (n^c_{0\beta}+1)\,P_\beta -\,n^c_{0\beta}\,P_0 \bigr]
+ {\color{blue} W_p(P_0 - P_1)}\,,
\end{aligned}\\[5pt]
&\begin{aligned}
\dot{P}_{1} &= \gamma_{01} \bigl[ {\color{blue} n^c_{01}}\,P_0 - ({\color{blue}n^c_{01}}+1)\,P_1\bigr] \\[2pt]
            &+ \gamma_{\alpha 1}\bigl[ n^c_{\alpha 1}\,P_\alpha -(n^c_{\alpha 1}+1)\,P_1 \bigr]
             + {\color{blue} W_p(P_1 - P_0)} \,,
\end{aligned}\\[5pt]
&\dot{P}_{\alpha} =\gamma_{\alpha 1}\bigl[(n^c_{\alpha 1}+1)\,P_1 -n^c_{\alpha 1}\,P_\alpha\bigr]
-(1+\chi)\Gamma\,P_\alpha \,,\\[5pt]
&\dot{P}_{\beta} = \gamma_{0\beta}\bigl[n^c_{0\beta}\,P_0 -(n^c_{0\beta}+1)\,P_\beta\bigr] + \Gamma\,P_\beta \,.
\end{align}
\end{subequations}

Note that the mean photon number $n^h_{01}$ of the hot reservoir in Eq.~(\ref{PE1_old}) is replaced by 
$n^c_{01}$ of the cold reservoir and the pumping term $W_p$ in Eq.~(\ref{PE1_new}). The mean photon number 
$n^h_{01} = 60,000$ of the previous 
studies~\cite{Scully2011,Svidzinsky2011,Dorfman2013,Creatore2013,Zhang2015,Killoran2015,Fruchtman2016}
corresponds to $W_p/\gamma_{01} \simeq 60,000$ and $W_p \simeq 1.1\times10^{15} {\;\rm s^{-1}}$.
It is instructive to compare this pumping rate with the number of photons incident per 
unit area per unit time for the black-body radiation of the Sun at temperature $T_s = 6000 {\;\rm K}$,
using the Planck distribution. The number of photons with energy greater than the energy gap $E_g = h\nu_g$ 
absorbed by the donor per unit area per unit time is given by
\begin{align}
Q_s(\nu_g,T_s) &= \frac{2\pi}{c^2}\int_{\nu_g}^\infty \frac{\nu^2}{e^{h\nu/k_BT_s} -1} d\nu  \,.
\end{align}
For $E_g=1.8{\;\rm eV}$, one obtains $Q_s\simeq 9.0\times 10^{25} {\;\rm m^{-2}\,s^{-1}}$. 
So the pumping rate $W_p= 10^{15}\,{\rm s}^{-1}$ corresponds to the photon flux
incident on a photocell with area $0.1\;\mu{\rm m}^2$.

\begin{figure}[htbp]
\includegraphics[width=1.0\linewidth]{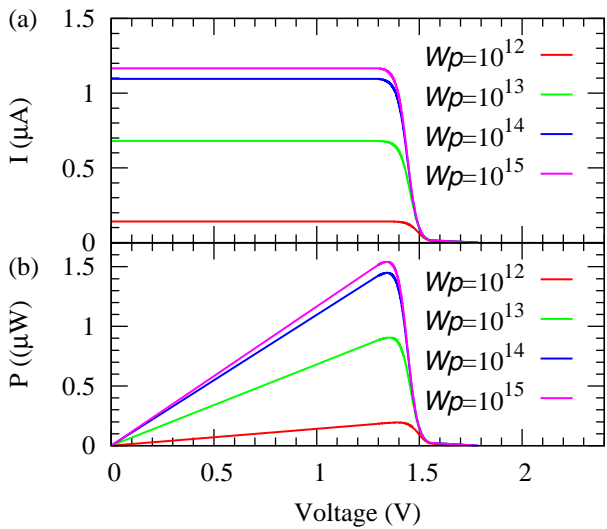}
\caption{(a) Current $I$ and (b) power $P$ are plotted as a function of voltage $V$ for different pumping rates $W_p$.}
\label{Fig2}
\end{figure}

Eq.~(\ref{PE1_new}) is solved numerically using the Runge-Kutta method. After the populations reach the steady state, 
the current is calculated as $I= e\Gamma P_\alpha$, and the voltage as 
$V=E_\alpha -E_\beta + k_BT\ln(P_\alpha/P_\beta)$. By changing the resistance $\Gamma$ of the external load from 
zero to infinity, one obtains the current-voltage curve of the photocell for various pumping rates, as shown in 
Fig.~\ref{Fig2}. The magnitude of current $I$ is readily estimated as follows. The generation rate of the excited 
electrons is propotional to the pumping rate, for example $W_p = 10^{12}\, {\rm s}^{-1}$. The transfer rate of 
the excited electrons to the acceptor is fast, i.e., the order of femtosecond. Thus, the current is just given by 
the product of electron charge and the generation rate, 
$I\sim 1.6\times 10^{-19}\,{\rm C}\times 10^{12}\,{\rm s^{-1}} = 0.16\,\mu{\rm A}$, i.e., the order of microampere.

\begin{figure}[htbp]
\includegraphics[width=1.0\linewidth]{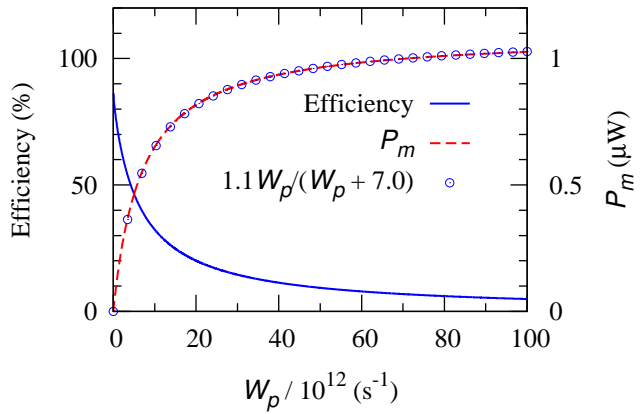}
\caption{Efficiency (blue solid line) and the maximum power $P_{\rm max}$ (red dashed line) are plotted as a function of 
the pumping rate $W_p$.}
\label{Fig3}
\end{figure}

We investigate how the efficiency and the maximum power change as a function of the pumping rate.
The efficiency $\eta$ of the photocell is calculated as
\begin {align}
\eta = \frac{P_{\rm out}}{P_{\rm in}} = \frac{P_m \;[{\rm\mu eV}]}{1.8{\;[\rm eV]}\cdot W_p\;[\rm s^{-1}]} \,.
\end{align}
It would be expected that the more power the photocell generates, the higher solar irradiance it receives. However,
as shown in Fig.~\ref{Fig3}, the maximum power increases linearly at the beginning but becomes saturated above 
a certain value of the pumping rate.  This implies that there is a bottleneck in population dynamics. 
A saturation curve like Fig.~{\ref{Fig3}} can be found in photosynthesis, which is well 
known as the photosynthesis-irradiance curve~\cite{Jassby76,Platt76,Jones2014}. 
Note that the maximum power as a function of pumping rate $W_p$ can be fitted by 
$P(W_p) = a\cdot W_p/(W_p + b)$ with $a =1.1$ and $b=7$. 
We find that the efficiency of the photocell decreases as the pumping rate $W_p$ increases.

\paragraph*{Photocell with coupled donors and an acceptor.---}
\label{sec:physics}

\begin{figure}[htbp]
\begin{tikzpicture}[thick, scale=1.0, 
>=stealth', pos=.8, photon/.style={decorate,decoration={snake,post length=1mm}} ]
\draw[] (-2.5,3.4) node {Hot reservoir};
\draw[] ( 2,3.4) node {Cold reservoir};
\draw [fill=red, opacity=0.50] (-3.5,-1.0) rectangle  (-2,3);
\draw [fill=green, opacity=0.20] (-1,-1.0) rectangle  (4.8,3);
\draw[color=red,->, photon]  (-2,0.6) -- (0.0,0.6);
\draw[blue,very thick]  (0,0)   node[left]{\color{black} $\ket{0}$} -- (2.0,0);
\draw[gray,very thick]  (0,2.0) node[left]{\color{black} $\ket{1}$} -- (2.0,2.0);
\draw[red, very thick]  (0,2.5) node[left]{\color{black} $\ket{2}$} -- (2.0,2.5);
\draw[black,thick, <->]  (0.2,0.04) -- node[midway,left] {$W_p$} (0.2,2.49);
\draw[black,thin, <->]  (1.0,0.04) -- node[midway, left] {$\gamma_{02}$} (1.0,2.49);
\draw[black,thin,<->] (1.5,0.04)  -- node[midway]       {$\times$} (1.5,1.99);
\draw[black,thin,<->] (1.5,2.04)  -- node[midway,right] {$\gamma_{12}$} (1.5,2.49);
\draw[black,thin, <->]  (2.0,2.0)-- node[below] {$\gamma_{1\alpha}$} (3,1.5);
\draw[black,thin, <->]  (2.0,2.5)-- node[midway] {$+$}  (3.2,1.5);
\draw[orange,line width=1.5pt] (3.0,1.5) -- (4.0,1.5) node[right] {\color{black}$\ket{\alpha}$};
\draw[cyan,  line width=1.5pt] (3.0,0.5) -- (4.0,0.5) node[right] {\color{black}$\ket{\beta}$};
\draw[black,thick, ->]   (3.5,1.5)  --  (3.5,0.5) node[midway,right] {$\Gamma$};
\draw[black, <-]  (1.7,0.04) -- (3.5,1.5) node[midway,left] {$\chi\Gamma$};
\draw[black,thin,<->] (2,0.04)   -- (3,0.48) node[midway,below right] {$\gamma_{0\alpha}$};
\draw[] (0.8,-0.75) node {Coupled donors};
\draw[] (3.6,-0.75) node {Acceptor};
\end{tikzpicture}
\caption{A photocell with coupled donors and an acceptor is in thermal equilibrium only with the cold reservoir. The
coupled donors form the bright state $\ket{2}$ and the dark state $\ket{1}$.}
\label{Fig4}
\end{figure}
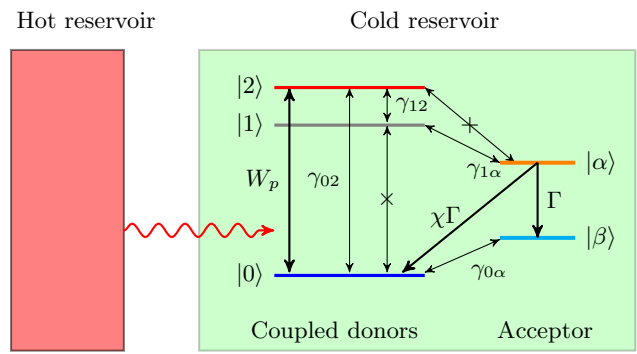

Let us turn to the main question how much the efficiency of a photocell is enhanced by quantum coherence.
Similar to the previous studies~\cite{Creatore2013,Zhang2015}, we consider the photocell model composed of 
two coupled donors and an acceptor, as shown in Fig.~(\ref{Fig4}). Unlike the previous studies, the photocell 
is in thermal equilibrium only with the cold reservoir. The dark state formed by the coupled donors plays 
a key role in enhancing the power of the photocell in compared with a photocell with uncoupled donors. 
The dynamics for occupation probabilities of energy level $E_i$ is readily given by the Pauli master equation 
\begin{subequations}
\begin{align}
&\begin{aligned}
\dot{P}_{0} &= \gamma_{01}\bigl[(1+{\color{blue} n^{c}_{01}})\,P_{1} - {\color{blue} n^{c}_{01}}\,P_{0}\bigr] 
           + \chi\Gamma\,P_{\alpha} \\[2pt]
            &+ \gamma_{0\beta}\bigl[(1+n^c_{0\beta})\,P_\beta -n^c_{0\beta}\,P_0\bigr] + {\color{blue} W_p(P_1 -P_0)}\,,
\end{aligned}\\[5pt]
&\begin{aligned}
\dot{P}_{1} &= \gamma_{01}\bigl[ {\color{blue} n^c_{01}}\,P_0 -(1+ {\color{blue} n^c_{01}})\,P_{1} \bigr] \\[2pt]
            &+ \gamma_{12}\bigl[ n^c_{12}\,P_2 -(1+ n^c_{12})\,P_{1}\bigr]+ {\color{blue} W_p(P_0 -P_1)}\,,
\end{aligned}\\[5pt]
&\begin{aligned}
\dot{P}_{2} &= \gamma_{12}\bigl[(1+n^c_{12})\,P_{1} - n^c_{12}\,P_{2} \bigr]\\[2pt]
            &+ \gamma_{\alpha 2}\bigl[n^c_{\alpha 2}\,P_{\alpha} -(1+n^c_{\alpha 2})\, P_{2}\bigr]\,,
\end{aligned}\\[5pt]
&\dot{P}_{\alpha}
=\gamma_{\alpha2}[(1+n^c_{\alpha2})\,P_{2} -n^c_{\alpha2}\,P_{\alpha}] -\Gamma(1+\chi)\,P_{\alpha}\,,\\[5pt]
&\dot{P}_{\beta}
=\Gamma P_{\alpha} + \gamma_{\beta 0}\bigl[ n^c_{\beta0}\,P_0 - (1+n^c_{\beta0})\,P_{\beta}\bigr]\,.
\end{align}
\label{Eq:PE_2donors}
\end{subequations}

\begin{figure}[htbp]
\includegraphics[width=0.95\linewidth]{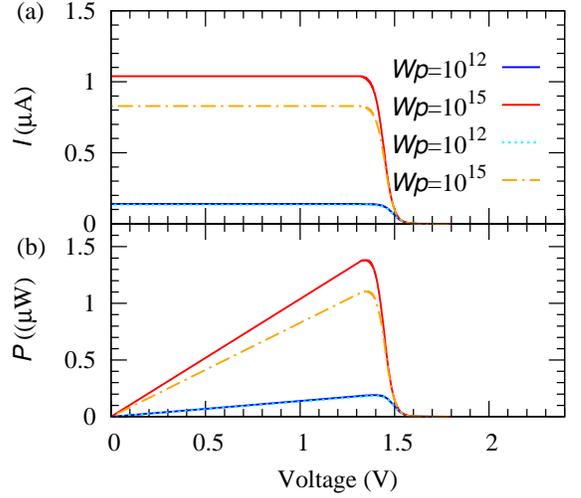}
\caption{(a) Current $I$ and power $P$ are plotted as a function of voltage $V$ for different pumping rates $W_p$,
and for the photocells with coupled donors (solid lines) and uncoupled donors (dashed and dash-dotted lines).}
\label{Fig5}
\end{figure}

\begin{figure}[htbp]
\includegraphics[width=0.95\linewidth]{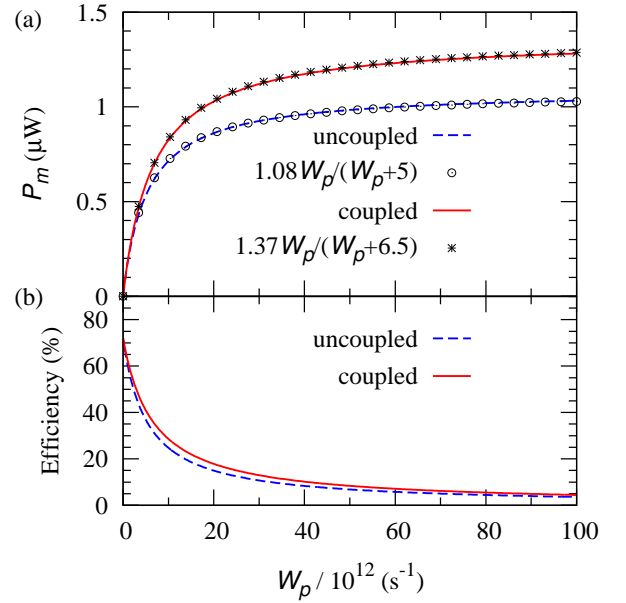}
\caption{(a) Maximum power $P_m$ and (b) efficiency as a function of the pumping rate $W_p$ for uncoupled donors 
(dashed lines) and coupled donors (solid lines). The maximum power curves for uncoupled and coupled donors are fitted by  
two functions $1.08 W_p/(W_p +5)$ and $1.37W_p/(W_p + 6.5)$, respectively.}
\label{Fig6}
\end{figure}

We solve Eq.~(\ref{Eq:PE_2donors}) with the parameters $\gamma_{ij}$ given by Refs.~\cite{Creatore2013,Zhang2015}.
We obtain the current-voltage curve and the power-voltage curve for different pumping rates and for photocells with 
uncoupled and coupled donors to see the effect of quantum coherence, as shown in Fig.~\ref{Fig5}. It is interesting 
that at low pumping rate $W_p = 10^{12}\,{\rm s}^{-1}$, the short-circuit current and the power are not enhanced
by the quantum coherence. However, at high pumping rate $W_p = 10^{15}\,{\rm s}^{-1}$, the quantum coherence gives 
rise to the strong enhancement of the short-circuit current and the power, agreeing with the previous
studies~\cite{Scully2011,Svidzinsky2011,Dorfman2013,Creatore2013,Zhang2015,Killoran2015,Fruchtman2016}.
Fig.~\ref{Fig6} depicts the maximum power $P_m$ and the efficiency as a function of pumping rate $W_p$ for coupled 
and uncoupled donors. For both uncoupled and coupled donors, the maximum power $P_m$ increases at first and becomes 
saturated as the pumping rate increases, but the efficiency decreases.
The photocell with coupled donors generates more power than that with uncoupled donors as the pumping rate increases,
but the enhancement in efficiency due to the quantum coherence is not the case. In contrast to the claim of the previous 
studies~\cite{Scully2011,Svidzinsky2011,Dorfman2013,Creatore2013,Zhang2015,Killoran2015,Fruchtman2016}, 
the enhancement of the efficiency due to the quantum coherence, via dark states or noise-induced quantum coherence,
is very small at $W_p=10^{15}\,{\rm s}^{-1}$ which corresponds to $n^h_{02} = 60,000$.

In conclusion, we proposed a new photocell model where the system is in thermal equilibrium only with the cold 
reservoir and the photon flux from the hot reservoir is described by the pumping term in the master equation. 
The pumping term resolves the problem of the incorrect mean photon number of the hot reservoir used by the 
previous studies. The maximum power and the efficiency were obtained as a function of the pumping rate. 
It is found that as the pumping rate increases, the power increases linearly at first but becomes saturated, 
and the efficiency 
decreases rapidly. It is shown that the quantum coherence via the dark state of the coupled donors clearly 
enhance the power significantly, but the efficiency tiny. Further study is needed to see whether quantum 
coherence could enhance the efficiency significantly and break the Shockley-Queisser limit of a single-junction 
solar cell. 

\vfill

\begin{thebibliography}{99}
%
\bibitem{Scovil59} H.E.D. Scovil and E.O. Schulz-DuBois, \prl\ {\bf 2}, 262 (1959).
\bibitem{Shockley1961} W. Shockley and H.J. Queisser, 
   {\it J. Appl. Phys.} {\bf 32}, 510 (1961).
\bibitem{Engel2007} G.S. Engel, T.R. Calhoun, E.L. Read. T.K. Ahn, T. Man\u{c}al, Y.-C. Cheng,
R. E. Blankenship, and G.R. Flemming, Nature {\bf 446}, 782 (2007).


%
\bibitem{Scully2010} M.O. Scully, \prl\ {\bf 104}, 207701 (2010).
\bibitem{Scully2011} M.O. Scully, K.R. Chapin, K.E. Dorfman, M.B. Kim, and A. Svidzinsky, 
   {\it Proc. Natl. Acad. Sci. USA} {\bf 108},  15097 (2011).
\bibitem{Svidzinsky2011} A.A. Svidzinsky, K.E. Dorfman, and M.O. Scully, 
   \pra\ {\bf 84}, 053818 (2011).
\bibitem{Dorfman2013} K.E. Dorfman, D.V. Voronine, S. Mukamel, and M.O. Scully, 
   {\it Proc. Natl. Acad. Sci. USA} {\bf 110}, 2746 (2013).
%
%
\bibitem{Creatore2013} C. Creatore, M.A. Parker, S. Emmott, and A.W. Chin,
   \prl\ {\bf 111}, 253601 (2013).
\bibitem{Zhang2015} Y. Zhang, S. Oh, F.H. Alharbi, G.S. Engel, and S. Kais,
   Phys. Chem. Chem. Phys. {\bf 17}, 5743 (2015).
\bibitem{Killoran2015} N. Killoran, S.F. Huelga, and M.B. Plenio,
   J. Chem. Phys. {\bf 143}, 155102 (2015).
\bibitem{Fruchtman2016} A. Fruchtman, R. G\'{o}mez-Bombarelli, B.W. Lovett, and E.M. Gauger
   {\it },
   \prl\ {\bf 117}, 203603 (2016).
%
\bibitem{Kirk2010} A.P. Kirk, \prl\ {\bf 106}, 0703 (2011).
\bibitem{Kirk2012} A.P. Kirk, Physica B {\bf 407}, 544 (2012); {\it ibid} {\bf 423}, 58 (2013).
\bibitem{Kirk2015} A.P. Kirk, J. Appl. Phys. {\bf 118}, 034506 (2015).
%
\bibitem{Mandel} L. Mandel and E. Wolf, {\it Optical coherence and quantum optics} (Cambridge University Press, New York, 1995)
\bibitem{Slichter} C.P. Slichter, {\it Principles of Magnetic Resonance} 3rd Ed., p 254 (Springer-Verlag, New York, 1989)
%

\bibitem{Jassby76}
A.D. Jassby and T. Platt, Limnol. Oceanogr. {\bf 21}, 540–547 (1976).
\bibitem{Platt76} T. Platt and A.D. Jassby, J. Phycol. {\bf 12}, 421 (1976).
\bibitem{Jones2014} C.T. Jones, S.E. Craig, A.B. Barnett, H.L. MacIntyre, and J.J. Cullen,
    J. Phycol. {\bf 50}, 341 (2014)
%
\end{thebibliography}
\end{document}